\documentclass[12pt]{article}

\title{\bf 
Phase Transition Strength through Densities\\
of General Distributions of Zeroes
}
\author{ 
{\it W. Janke}\\
Institut f\"ur Theoretische Physik,\\
Universit\"at Leipzig,\\
Augustusplatz 10/11, \\
04109 Leipzig, Germany \\
{}\\
{\it D.A. Johnston}\\ 
Department of Mathematics,\\
Heriot-Watt University,\\
Riccarton,\\
Edinburgh, EH14 4AS, Scotland
\\
{\bf and}
\\
{\em R. Kenna}\\
School of Mathematical and Information Sciences,\\
Coventry University,\\
Coventry, CV1 5FB, England
}
\begin{document}
\maketitle
                      {\Large
                      \begin{abstract}
%
A recently developed technique for the determination of the density of 
partition function zeroes using data coming from finite-size systems 
is extended to deal with cases where  the zeroes are 
not restricted to a curve in the complex 
plane and/or come in degenerate sets. 
The efficacy of the approach is demonstrated by application to
a number of models for which these features are manifest and the zeroes
are readily calculable.
%
                        \end{abstract} }
%
  \thispagestyle{empty}
%
%
  \newpage
%
                  \pagenumbering{arabic}

\section{Introduction}
\setcounter{equation}{0}

The study of phase transitions is central to statistical mechanics. 
Of primary interest is the determination of the location, the order 
and the strength of the transitions. 
While only systems of infinite extent display 
such phenomena, these are not directly accessible to the non-perturbative 
computational approach, which is restricted to a finite number of 
degrees of freedom. 
There are, however, well-established techniques for the extraction 
of information from numerical studies of finite systems, and 
prominent amongst them is finite-size scaling (FSS).

The FSS hypothesis is based on the premise that the
only relevant scales are the correlation length of the infinite-size
system and the linear extent of its finite-size counterpart 
\cite{FSS}.
A modification, in which the 
correlation length of the finite system replaces its actual size,
extends the validity of the hypothesis to the upper critical dimension  
\cite{KeLa}.
Traditional techniques to determine phase transition strength
from finite-size data involve the application of FSS to 
thermodynamic quantities or to the lowest lying partition function 
zeroes \cite{IPZ}.

However, a full understanding of the properties of the infinite-size
system requires knowledge of the density of zeroes too. 
While it has
long been expected that extraction
of this quantity from finite-size systems
would be a lucrative source of information, 
a technique to do so proved elusive \cite{Martin}. 
The source of the 
difficulties is that it involves reconstruction of a continuous density
function from a discrete data set, or sets, as the density of zeroes
for a finite system is essentially a set of delta functions.
Recent considerations have bypassed these difficulties 
\cite{usFisher,usLY}.
Rather than focusing on the density of zeroes itself, one determines
the integrated density of zeroes. The robustness and efficiency of this
approach was demonstrated in \cite{usFisher,usLY} 
and the method favourably compared
to other techniques in \cite{AlHa02}.

In these previous analyses, the distribution of zeroes had two special 
properties. These are ({\em i\/}) the zeroes dominating critical or 
pseudocritical behaviour lie on a curve 
called the {\emph{singular line}}, which impacts onto the real axis at the
transition point  and ({\em ii\/}) these zeroes are simple zeroes (zeroes
of order one).
While these two properties are common to the bulk of models in statistical 
physics
and in lattice field theory, they are by no means generic and
the question of the generality of the technique presented  
in \cite{usFisher,usLY}
therefore arises. 

The purpose of this paper is to
extend the method presented in
\cite{usFisher,usLY} to deal with situations where the 
above two properties do 
not hold. 
Instead, the method developed here in Sec.~2 assumes the zeroes to be distributed 
across a two-dimensional
region in the
complex plane and/or to occur in degenerate sets. Such distributions
of zeroes have been observed in various models of statistical physics and 
lattice field theory in two dimensions. The models we
address in Sec.~3 are  
(a) the Ising model on a square lattice 
(using Brascamp-Kunz boundary conditions)
with anisotropic couplings,
(b) the Ising model on a 
bathroom-tile lattice, and 
(c) the case of free Wilson fermions in two dimensions. 
While all of these models are in the same two-dimensional Ising universality 
class,
their detailed distributions of zeroes are quite different and provide a 
sufficiently
wide sample to test the improved density-of-zeroes approach to the 
detection and characterization of phase transitions. 
Finally, Sec.~4 contains our conclusions.

\section{Zeroes and their Densities}
\setcounter{equation}{0}

All of the information on a thermodynamical system 
in equilibrium is encoded in the
zeroes of the appropriate partition function. Indeed, for a system
of finite size, when the partition function, $Z_L$, can be 
written in as
a polynomial in an appropriate function, $z$, of temperature, field
or of a coupling parameter, we may write
\begin{equation}
 Z_L(z) = A(z) \prod_j{\left( z-z_j(L) \right)}
\quad ,
\label{Z}
\end{equation}
where $L$ denotes the linear extent of the system, $j$ labels the zeroes,
and $A(z)$
is a smooth non-vanishing function which plays no crucial 
role in the sequel and is henceforth discarded.
 
In numerical approaches to critical phenomena, FSS of the zeroes,
$z_j(L)$ (with $j$ fixed -- typically to $j=1$, which labels the 
zero nearest the transition point),
is used to determine properties of phase transitions. 
A summary of the 
status of some of these calculations is given in \cite{usFisher}. 
On the other hand, attempts have also been made to gain a deeper 
understanding
of some more tractable models analytically \cite{Abe,SuzukiLY}. 
Where these attempts have
involved zeroes of the partition function, it is clear that much 
information is contained in their density. 
The technique developed in \cite{usFisher} 
is essentially a convergence
of these two approaches, and we summarize it here for
convenience.

\subsection{Simple Zeroes on a Singular Line}
The reduced free energy is obtained from (\ref{Z})
as
\begin{equation}
 f_L(z) = \frac{1}{V} \ln{Z_L(z)}
 =
 \frac{1}{V} \sum_j{\ln{\left( z - z_j(L) \right)}}
\quad ,
\label{f}
\end{equation}
having discarded the regular contribution coming from $A(z)$.
Here $V$ represents the volume of the system.
In independent series of publications, Abe \cite{Abe}
and Suzuki \cite{SuzukiLY} 
assumed that the zeroes fall on a singular line in the
complex plane, parameterized by $z=z_c+r \exp(i \phi)$, where $z_c$
is the transition point. 
In this case, a necessary and sufficient condition 
to achieve the correct scaling behaviour for the specific heat is 
that the density of zeroes along the singular line in the thermodynamic
limit behave as
$g_\infty{(r)} \propto
 r^{1-\alpha}$,
where $\alpha$ is the usual critical exponent of the specific heat.
Integrating, gives the cumulative density of zeroes in the infinite-volume
limit,
\begin{equation}
 G_\infty{(r)}
 \propto
 r^{2-\alpha}
\quad .
\label{Gal}
\end{equation}
In the finite-volume case, the density of zeroes is a string of delta functions, and,
\begin{equation}
 g_L{(r)}
 =
\frac{1}{V} \sum_j{\delta(r-r_j(L))}
\quad ,
\end{equation}
where the $j^{\rm{th}}$ zero is given by $z_j(L) = z_c + r_j(L)\exp(i \phi)$.
Integrating
this along the singular line 
leads to the following expression for the cumulative density of zeroes
 \cite{usFisher,usLY}:
\begin{equation}
G_L(r) = 
\left\{
        \begin{array}{cl}
              j/V & \mbox{if $r \in (r_j,r_{j+1})$} \quad, \\
             (2j-1)/2V & \mbox{if $r =r_j$}      \quad.
        \end{array}
\right.
\label{sens}
\end{equation}

The two central observations of \cite{usFisher,usLY} 
were, firstly, that equating the infinite-volume
density formula (\ref{Gal}) 
to its finite-volume counterpart (\ref{sens}) 
is sufficient (with hyperscaling)
to recover standard FSS expressions (indeed, FSS, traditionally 
the consequence of a hypothesis, 
emerges quite naturally from this approach), 
and, secondly, that (\ref{sens})
is a sensible definition of
the cumulative density of zeroes in the finite case. With this 
definition, the strength of transitions may be {\emph{directly}} 
measured by fitting  to the
ansatz
\begin{equation}
 G(r) = a_1 r^{a_2} + a_3
\quad .
\label{2nd}
\end{equation}
In particular, a non-zero value of $a_3$ indicates a definite phase. When 
$a_3$ vanishes,
a transition of first order is indicated if $a_2 \sim 1$, while a value of
$a_2$ larger than $1$
indicates a second-order transition with strength
\begin{equation}
 \alpha = 2 - a_2
\quad .
\label{aa2}
\end{equation}
In \cite{usFisher} and \cite{usLY} 
this method was tested by application to a number of models
in statistical physics and in lattice field theory. 
In all of these models,
the locus of zeroes is one-dimensional, 
with a singular line impacting on the
real axis at the transition point. Furthermore, all zeroes
for finite lattices were simple zeroes (with no degeneracies).
The question now arises as to how the technique translates to 
more general distributions of zeroes.

\subsection{General Distribution of Zeroes}
Departures from such smooth linear sets of zeroes were first
observed for models on
 hierarchical and anisotropic two-dimensional 
lattices, for which there can exist 
a two-dimensional distribution (area) of zeroes 
\cite{More2D,SaKu84,StCo84}. 
Since then, a host of systems have been discovered with this 
feature \cite{Moremore2D,JaJo02,KeSe02,ChHu03}.
A common characteristic of all such
two-dimensional distributions of zeroes is that the only
 physically 
relevant point at which they cross the real axis,
  in the thermodynamic limit,
is that which corresponds to the phase transition. 
It is, however, possible that the zeroes cross the real or imaginary
axis at unphysical points. These points may be associated with new
universality classes.

Stephenson \cite{St87} has shown that  the density of zeroes 
for such two-dimensional distributions
in the infinite-volume limit is
\begin{equation}
 g_\infty(x,y) = y^{1-\alpha-m} f\left( \frac{x}{y^m}\right)
\quad,
\label{2Dinfinite}
\end{equation}
where $(x,y)$ give the location of zeroes in the complex plane, with the
critical point as the origin. 
Here $m$ is a new type of exponent
which is related to the shape of the two-dimensional distribution \cite{St87}. 

Integrating out the $x$-dependence in 
(\ref{2Dinfinite}) yields \cite{St87}
\begin{equation}
 g_\infty(y) = \int_{x_1}^{x_2}{g_\infty(x,y) dx}
 \propto y^{1-\alpha}
\quad ,
\end{equation}
where $x_1$ and $x_2$ mark the extremities of the distribution
of zeroes at a distance $y$ from the $x$ axis in the complex plane.
Integrating again, to determine the cumulative density of zeroes
at the point $r$ in the  $y$-direction, yields
\begin{equation}
 G_\infty{(r)}
 \propto
 r^{2-\alpha}
\quad ,
\end{equation}
an expression identical to (\ref{Gal}).
The strength of the transition, as measured by $\alpha$,
can therefore be determined by similar methods to those 
previously used.
However, rather than counting the zeroes along the singular line, 
one now counts
them up to a line $y=r$ within the two-dimensional complex domain 
they inhabit.

\begin{figure}[t]
\vspace{5cm}
\includegraphics{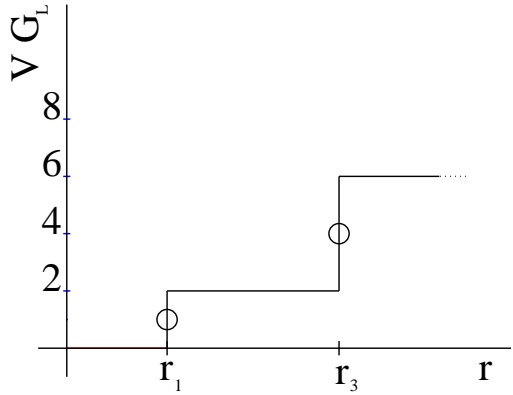}
\caption[a]{Schematic plot of cumulative density of zeroes
as defined by (\ref{soln}). In this example, where the volume, $V$, is
fixed, $r_1=r_2$
are $2$-fold degenerate, while $r_3=\dots = r_6$ are
$4$-fold degenerate. }
\label{Des2}
\end{figure}

The second new feature we wish to accommodate is the existence
of degeneracies in the set of zeroes. If a number of zeroes
coincide, $G_L$, as defined in (\ref{sens}), is multivalued
and is no longer a proper function. 
A more appropriate density function is determined
as follows.
Suppose, in general, that $z_j=z_{j+1}= \dots 
= z_{j+n_j-1}$ 
are $n_j$-fold
degenerate. 
By a glance at Fig.~\ref{Des2}
it is easy to convince oneself
that the densities to the left and 
right of
an actual zero are given by
\begin{equation}
V G_L(r) = \left\{
                \begin{array}{ll}
                j+n_j-1 & \mbox{for $r \in (r_{j+n_j-1},r_{j+n_j})$
 \quad ,} \\
                j-1     & \mbox{for $r \in (r_{j-1}  ,r_j    )$ \quad .}
                \end{array}
         \right.
\label{lr}
\end{equation}
The density at the $n_j$-fold degenerate zero, $r_j$, is again 
sensibly defined as an average:
\begin{equation}
 G_L(r_j) = \frac{1}{V}
                 \left(
                        j+\frac{n_j}{2}-1
                 \right)   
\label{soln}
\quad .           
\end{equation}
This is the most general formula for extracting the density
of any distribution of zeroes and deals with two-dimensional 
spreads and degeneracies. Fitting this quantity to the form
(\ref{2nd}) yields the strength of a second-order transition
through (\ref{aa2}).
As in \cite{usFisher} and \cite{usLY}, the criteria for a good
fit are good data collapse in $L$ (or $V$) 
and $j$ near the transition point
and $a_3$ be compatible with zero.

The error estimates appropriate to this modified density analysis
may be determined from a procedure
adapted from \cite{usFisher} and which we now elucidate.
In the present case, where zeroes may be degenerate, 
the monotone nature of the cumulative 
density function means that the actual value of 
$G_L(r_j(L))$ cannot deviate from   (\ref{soln}) by more than
$\pm n_j/2V$ (see Fig.~\ref{Des2}). 
The quantitative difference between this
starting point and that in \cite{usFisher} is that
this deviation is not constant in this case.

Let $G_j^{\rm{obs}}(L)$ represent the data point coming from the 
size-$L$ lattice and corresponding 
to the $j^{\rm{th}}$ zero, which is $n_j$-fold degenerate. 
Assign an initial
error $\sigma_j(L) = \sigma_{\rm{arb}}n_j/V$ to this data point,
where $\sigma_{\rm{arb}}$ is arbitrary. 
With these errors, the appropriate goodness-of-fit is given by
\begin{equation}
 \chi_1^2
 =
 \sum_{L,j}{
            \frac{
                  [G_j^{\rm{obs}}(L)
                  -G_j^{\rm{exp}}(L)
                  ]^2
                 }{
                  \sigma_j(L)^2
                 }
           }
 =
 \sum_{L,j}{
            \frac{V^{2}}{
                          \sigma^2_{\rm{arb}} n_j^2
                         }
           [G_j^{\rm{obs}}(L)
           -G_j^{\rm{exp}}(L)
           ]^2
           }
\quad ,
\label{minchi1}
\end{equation}
where  the expected density value, $G_j^{\rm{exp}}(L)$, comes from 
the model (\ref{2nd}). 
Minimizing $\chi_1^2$ yields 
 the parameters $a_i$ in 
(\ref{2nd})
with associated 
 errors  denoted $d a_i^{\rm{arb}}$.

Assume, now, the actual error associated with each data point is, 
in fact, $\sigma n_j/V$. 
The corresponding chi-squared may be written \cite{usFisher}
\begin{equation}
 \chi_2^2
 = 
 \frac{\sigma^2_{\rm{arb}}}{\sigma^2}\chi_1^2
\quad .
\end{equation}
If the model fits well, $\chi_2^2/N_{\rm{dof}}$ should be close to unity, 
where $N_{\rm{dof}}$ is the number of 
degrees of freedom. The error assigned to each
point now becomes
\begin{equation}
 \sigma^2
 =
 \sigma_{\rm{arb}}^2 \chi_1^2/N_{\rm{dof}}
 =
 \chi_1^2/N_{\rm{dof}}
\quad ,
\end{equation}
having chosen $ \sigma_{\rm{arb}}$ to be unity.
Moreover, the actual errors associated with the parameters
$a_i$ are (with $ \sigma_{\rm{arb}}=1$) 
\begin{equation}
da_i =   \frac{\sigma}{\sigma_{\rm{arb}}} d a_i^{\rm{arb}}
     =   \sigma  d a_i^{\rm{arb}}
\quad .
\end{equation}
Just as in \cite{usFisher}, this approach
prohibits an independent goodness-of-fit test.

In summary, the proceedure is to let 
 $\sigma_j(L)= n_j/V$ and minimise
$\chi_1^2$ in (\ref{minchi1})
to find $a_i$ and $d a_i^{\rm{arb}}$. The best estimates
for the errors are, then,
$
 da_i = \sqrt{\chi_1^2/N_{\rm{dof}}} \, d a_i^{\rm{arb}}
$.

Note that standard FSS is for fixed-index zeroes and gives that the
distance of a zero from the critical point is 
\begin{equation}
 r_j(L) \sim L^{-1/\nu}
\quad .
\label{IFSS}
\end{equation}
Typically one uses the imaginary part of the zero, ${\rm{Im}} z_j$, for
the distance $r_j$ in a traditional FSS analysis.
The real part of the lowest zero may be considered as a pseudocritical point.
Its scaling is characterised by the so-called shift exponent, $\lambda$, and
\begin{equation}
 {\rm{Re}}z_1(L) -z_c \sim L^{-\lambda}
\quad ,
\label{RFSS}
\end{equation}
where $z_c$ marks the critical point.
Usually $\lambda$ coincides with $1/\nu$, but this is not always the case
and the actual value of the shift exponent depends on the lattice
topology.
For a summary of some recent results concerning the finite-size shifting 
of the pseudocritical point in the Ising case in 
two dimensions, see \cite{JaKe02}.

\section{Testing the Method on Various (Ising) Models}
\setcounter{equation}{0}

We take three two-dimensional models for which the zeroes are easily
calculated. In each case the real, physical, critical point is in the
Ising universality class, with strength of transition given
by $\alpha = 0$ (corresponding to a logarithmic divergence in the
specific heat).

\subsection{Square Lattice Ising Model with Anisotropic Couplings}
%

The task of analytically solving the Ising model in two dimensions for
finite-size systems is greatly ameliorated by the usage of 
Brascamp-Kunz boundary conditions \cite{JaKe02,BK},
where for an $M \times 2N$ lattice,
the spins in the left boundary row at $x=0$ are fixed to $+$ and 
in the right boundary row at $x=M+1$ to the alternating sequence $+-+-\dots$,
whereas in the $y$-direction periodic boundary conditions are assumed. 
In the general case of anisotropic couplings -- $J_1$ along the $x$- and
$J_2$ along the $y$-direction, with arbitrary ratio 
$R = J_2/J_1$ -- the partition function takes the form \cite{boris}
\begin{equation}
Z_{M,2N} = 2^{2NM} \prod_{i=1}^M \prod_{j=1}^N
\left[
\cosh(2\beta) \cosh(2R\beta) - \sinh(2\beta) \cos(\phi_i)
- \sinh(2R\beta) \cos(\theta_j)
\right] \quad ,
\label{eq:BK_part}
\end{equation}
where $\phi_i = i\pi/(M + 1)$, $\theta_j = (2j - 1)\pi/2N$,
and $\beta = J_1/k_B T$.
Recall that for fully periodic boundary conditions, the analogue of 
(\ref{eq:BK_part}) consists of a sum of four product terms \cite{Kaufman}
which is much more cumbersome to analyze for the zeroes.

For isotropic couplings with $R=1$ the term in square brackets of (\ref{eq:BK_part})
simplifies to $1 - 2\xi \sinh(2\beta) +\sinh^2(2\beta)$, with 
$-1 \le \xi = (\cos{\phi_i} + \cos{\theta_j})/2 \le 1$.
It immediately follows that the complex zeroes can be parametrized exactly as
$\sinh(2\beta) = \xi \pm i \sqrt{1-\xi^2}$, i.e., that they are distributed on the unit 
circle in the complex $\sinh(2\beta)$-plane.

\begin{figure}[bth]
\vspace{6cm}
\includegraphics{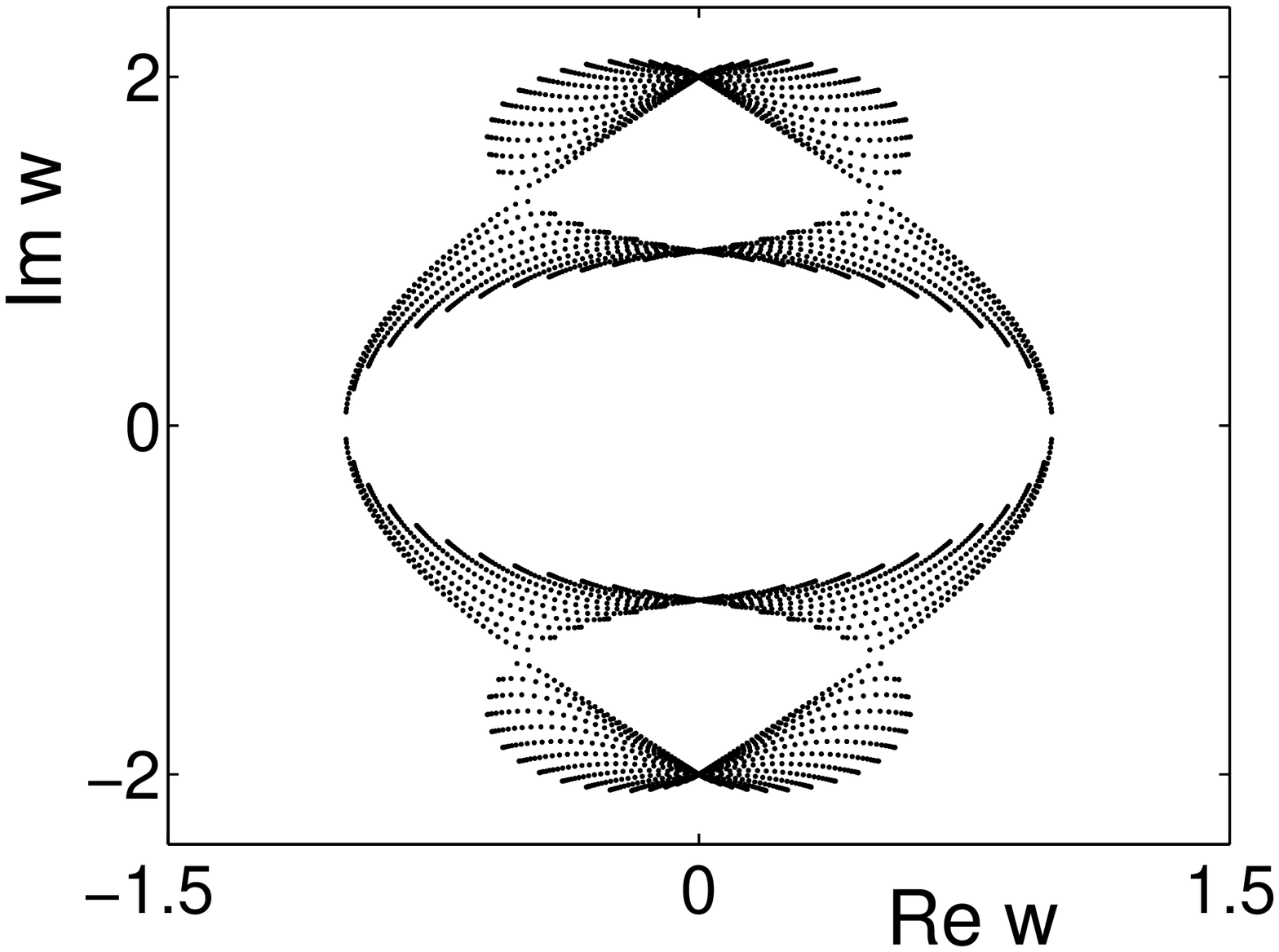}
\caption[a]{The partition function zeroes in the complex 
$w=2\sinh(2 \beta)$ plane for the anisotropic ($J_2 = 3 J_1$)
$L=M=2N=40$ Ising 
model with Brascamp-Kunz boundary conditions.}
\label{zeroesBK40}
\end{figure}

For the anisotropic model with $R=3$,
each factor in (\ref{eq:BK_part})
can be rewritten as a 
fourth-order polynomial in $w=2\sinh( 2 \beta)$
to give
\begin{equation}
Z_{M,2N} =  \prod_{i=1}^M \prod_{j=1}^N
\left[
w^4 + 5 w^2 + 4 -  2 w \cos(\phi_i)
-  ( 6 w + 2 w^3) \cos(\theta_j)
\right] \quad .
\label{eq:BK_part2}
\end{equation}
The zeroes of (\ref{eq:BK_part2}) are also easily determined numerically, 
but it is not possible to parametrize them by a single variable,
implying that they are distributed across a two-dimensional region rather 
than on a one-dimensional curve as in the isotropic case.
In Fig.~\ref{zeroesBK40} this two-dimensional distribution of zeroes
is shown for the case $R = 3$ 
and a square lattice of size $L= M = 2N = 40$.

The zeroes impact onto the real axis at the point $w=1$ and the
critical behaviour is expected to be dominated by the
zeroes close by. 
The zeroes in this case are all simple zeroes (no degeneracies),
so it should be noted that this case is essentially
a test of the applicability of the 
method to the situation of a two-dimensional
distribution of zeroes in the complex plane rather than a test
of how the method copes with varying degeneracies.

\begin{figure}[t]
\vspace{5cm}
\includegraphics{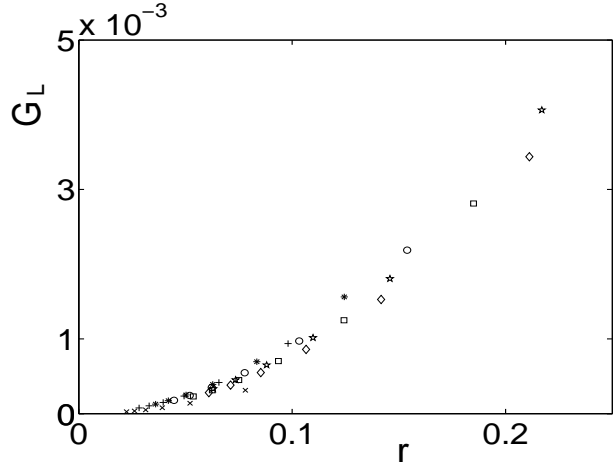}
\caption[a]{The distribution of zeroes near $w=1$ for the anisotropic
Ising model with anisotropy ratio $R=3$ subject to Brascamp-Kunz boundary conditions
for $L=40-140$ 
and  $j=1$ ($\times$), 
     $j=2$ ($+$),
     $j=3$ ($*$),
     $j=4$ ($\circ$),
     $j=5$ ($\put(1,0.5){\framebox(5,5)}~~$),
     $j=6$ ($\diamond$),
     $j=7$ ({\tiny{$\bullet$}}),
     $j=8$ ($\star$).
}
\label{GBK}
\end{figure}

The cumulative-density distribution for this set of zeroes
is plotted in Fig.~\ref{GBK}.
A 
three-parameter fit to (\ref{2nd}) for the first $8$ zeroes for 
lattices of size
$L=40,60,80,100,120$, and $140$ gives $a_3=0.000002(15)$, indicating the 
presence of a transition. With $a_3$ set to zero,
a two-parameter fit then yields $a_2 = 2.016(32)$,
close to the expected value of $2$ (which corresponds to 
$\alpha = 0$).

A closer inspection of Fig.~\ref{GBK} shows that
the $j=1$ zeroes
(denoted by the symbol $\times$) are slightly misaligned with respect 
to the higher-index zeroes. 
We have therefore repeated the fit 
restricted to $j=2-8$, which yields
$G(r) = 0.088(7) r^{2.008(33)}$, so that $\alpha = -0.008(33)$.
This is nice confirmation that the technique works when the distribution 
of (non-degenerate) zeroes is two-dimensional.

Standard FSS applied to fixed-index zeroes using (\ref{IFSS})
yields the expected result, $\nu = 1$ \cite{JaKe02}. 
Similarly, the  shift exponent in (\ref{RFSS}) is found to be
$ \lambda=2$. Thus $\lambda$ is not coincident with $1/\nu$.
This contrasts with the case of the Ising model in two dimensions with
toroidal boundary conditions \cite{FF} but matches
results using topologies with a trivial fundamental
 homotopy group \cite{homotopy0}. 

To understand these numerical results we return to the finite lattice expansion of
(\ref{eq:BK_part2})
and look at the finite-size scaling of the lowest zero $w_1$, which is given
by the roots of the factor in (\ref{eq:BK_part2}) with $i=j=1$
on an $M \times 2 N$ lattice. For an infinite lattice
the expression factorizes to give $(4 +w^2) ( 1 - w)^2$
and we see the points where the distribution pinches down as the roots
at  $w=1$ (and at $w = \pm 2i$).
For a finite square lattice
($L=M = 2 N$) we can expand around the root at
$1$ in powers of  ${1/L}$ to
find
\begin{equation}
w = 1 + { \pi i \over L} - { \pi ( 2 i + 5 \pi ) \over 10 L^2 } +
\cdots 
\quad .
\end{equation}
Separating the real and imaginary parts yields 
${\rm{Im}} w_1 (L) \sim L^{-1}$ and
${\rm{Re}} w_1 (L) - w_{c}  \sim  L^{ - 2}$.

\begin{figure}[t]
\vspace{5cm}
\includegraphics{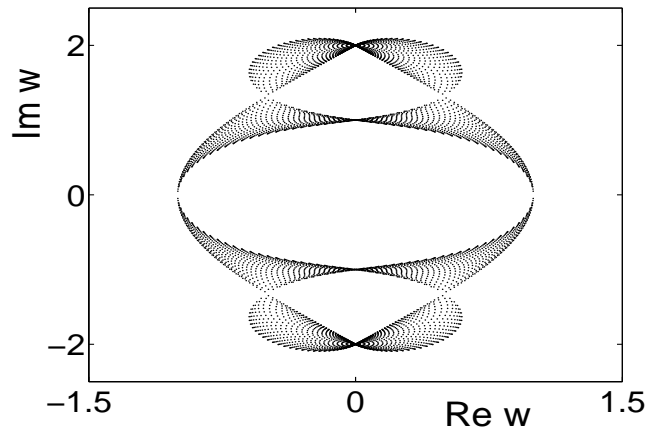}
\caption[a]{The partition function zeroes in the complex 
$w = 2 \sinh(2\beta)$ plane for the anisotropic ($J_2 = 2 J_1$)
$L=M=2N=40$ Ising 
model with Brascamp-Kunz boundary conditions.}
\label{zeroesBK40s=2}
\end{figure}

\begin{figure}[bht]
\vspace{6.5cm}
\includegraphics{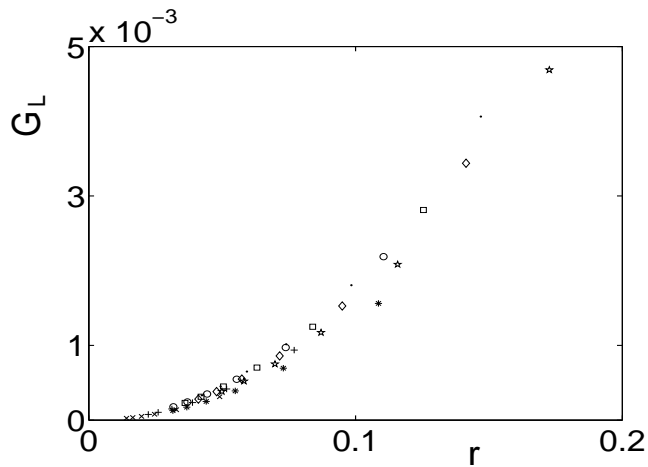}
\caption[a]{The distribution of zeroes near $w=1$ for  the 
Ising model with anisotropy ratio $R=2$ subject to Brascamp-Kunz boundary conditions
for $L=40-140$ 
and  $j=1$ ($\times$), 
     $j=2$ ($+$),
     $j=3$ ($*$),
     $j=4$ ($\circ$),
     $j=5$ ($\put(1,0.5){\framebox(5,5)}~~$),
     $j=6$ ($\diamond$),
     $j=7$ ({\tiny{$\bullet$}}),
     $j=8$ ($\star$).
}
\label{GBKs=2}
\end{figure}

For comparison,
we present a similar analysis with anisotropy ratio $R=2$, for which the 
zeroes are plotted in Fig.~\ref{zeroesBK40s=2}. While the overall
shape of the distribution is the same as in the $R=3$ case of
Fig.~\ref{zeroesBK40}, its detailed structure is different.
In this case the density analysis reveals $a_3=-0.000\,01(2)$, and a 
subsequent two-parameter fit to the first $8$ zeroes for 
lattices of size $L=40$--$140$  yields $a_2=2.009(30)$,
i.e.\ $\alpha = -0.009(30)$.
The corresponding data is displayed in Fig.~\ref{GBKs=2}.
%

\subsection{Bathroom-Tile Lattice}
\begin{figure}[htb]
\vspace{7cm}
\includegraphics{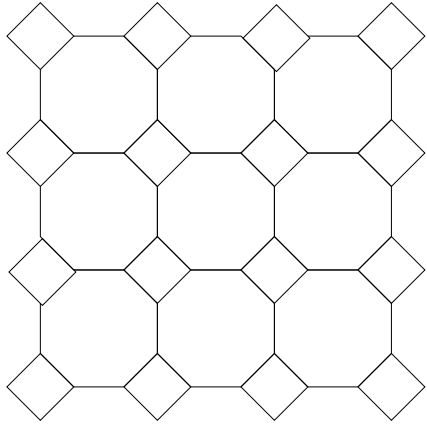}
\caption[a]{The bathroom tile lattice.}
\label{btile}
\end{figure}

%
It is also possible to obtain two-dimensional distributions of zeroes 
for two-dimensional Ising models with {\em isotropic} couplings,
one example being the Ising model on a bathroom-tile lattice \cite{Moremore2D}. 
This is the $(4 \cdot 8^2)$ lattice depicted in Fig.~\ref{btile} and 
which is dual 
to the Union Jack lattice.

\noindent
The continuum form of the (reduced) free energy on the  bathroom-tile lattice
is given by
\begin{eqnarray}
\label{fbath}
f &=& \frac{3}{2} + \frac{1}{2} \ln ( 1 + u)  \\
&+& \frac{1}{8} \int_{-\pi}^{\pi}   \int_{-\pi}^{\pi}{
{d \theta_1 d \theta_2 \over ( 2 \pi)^2} \ln \left[ A(u) +B(u) ( \cos
  (\theta_1)
+ \cos (\theta_2) ) + C(u) \cos(\theta_1) \cos ( \theta_2) \right]} \quad ,
\nonumber
\end{eqnarray}
where $u= \exp( - 2 \beta)$ and
\begin{eqnarray}
A(u) &=&  (1 + u^2)^2 ( 1 - 4 u + 10 u^2 - 4 u^3 + u^4) \quad , \nonumber \\
B(u) &=&   2 u ( 1 - u)^3 ( 1 + u ) ( 1 + u^2) \quad , \\
C(u) &=&   - 4 u^2 (1 - u)^4 \quad . \nonumber
\end{eqnarray}
This system is 
described in detail in \cite{Moremore2D}. The zeroes of the partition function
were calculated from the finite lattice discretization of 
{\it one} of the terms in the partition function for periodic boundary conditions leading to (\ref{fbath}), namely
\begin{eqnarray}
 Z &=& 2^{MN} \prod_{r=1}^{M} \prod_{s=1}^{N} \left\{ A(u) +B(u) \left[ \cos
   \left( { 2 r - 1 \over M }\right)  
 + \cos \left( {2 s - 1 \over N}\right) \right] \right. \nonumber \\ 
&+& \left.  C(u) \cos \left({2 r - 1 \over M} \right) 
 \cos \left( {2 s - 1 \over N} \right) \right\}^{1/2} \quad .
\label{Zbath}
\end{eqnarray}
In principle the full partition function is a sum of four%
\footnote{One of which will vanish at criticality for toroidal topology.}
such terms, 
differing in the arguments of the cosines
which correspond to the four possible choices of (anti)periodic boundary conditions for the 
two species of fermions in the continuum limit of the model. In using 
(\ref{Zbath}), we are assuming that the scaling behaviour
of one of these terms is generic. An alternative, which we do not pursue here as we are essentially
interested in testing the scaling of the cumulative density of zeroes
rather than formulating the finite lattice 
models themselves, would be to construct Brascamp-Kunz type boundary conditions for the bathroom tile 
lattice. This would also have the effect of projecting out a (different) single product term in 
the expression for $Z$.

\begin{figure}[t]
\vspace{6cm}
\includegraphics{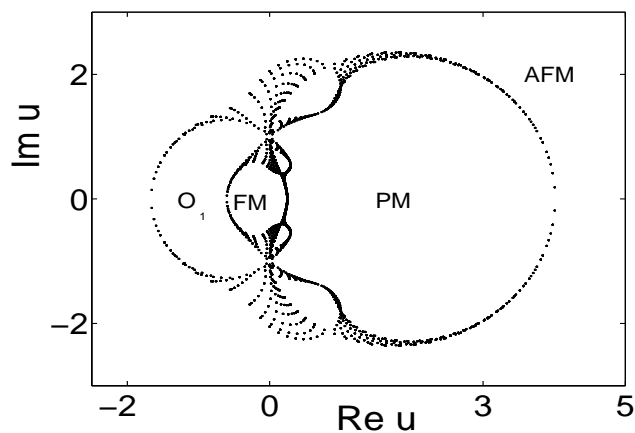}
\caption[a]{The partition function zeroes in the $u = \exp{(-2 \beta)}$
plane for the bathroom-tile Ising
model (\ref{Zbath}) with $L=M=N=40$. Here AFM, PM, FM and O$_1$ indicate the 
anti-ferromagnetic,
paramagnetic, ferromagnetic and unphysical phases, respectively.}
\label{zeroes_bath40}
\end{figure}

\begin{figure}[bh]
\vspace{6cm}
\includegraphics{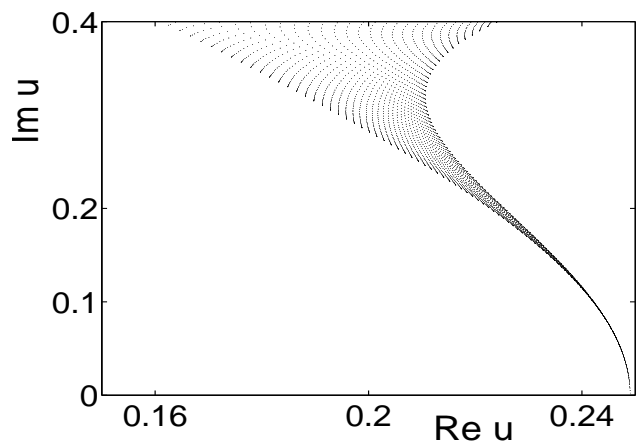}
\caption[a]{The bathroom-tile Ising zeroes
near the ferromagnetic critical point $u=0.249\,038\,4\dots$ for $L=M=N=200$. }
\label{2490384}
\end{figure}

The phase diagram for such a system has paramagnetic [PM],
ferromagnetic [FM] and anti-ferromagnetic [AFM] phases as well as an 
unphysical
phase which we denote as O$_1$, to adhere to the same notation as 
\cite{Moremore2D}.
The zeroes have varying degrees of degeneracy.
Those for $L=M=N=40$ are depicted in Fig.~\ref{zeroes_bath40}
in the complex $u = \exp{(-2 \beta)}$ plane 
and a blow-up of the region near the ferromagnetic critical point
for $L=200$ is given
in Fig.~\ref{2490384}. Zeroes in the vicinity of the critical point
taper off into a quasi-one-dimensional locus, so the bathroom-tile case
is a test of the applicability of the method to  zeroes of varying
degeneracies,
rather than to a true two-dimensional distribution. 

\begin{figure}[tb]
\vspace{6.0cm}
\includegraphics{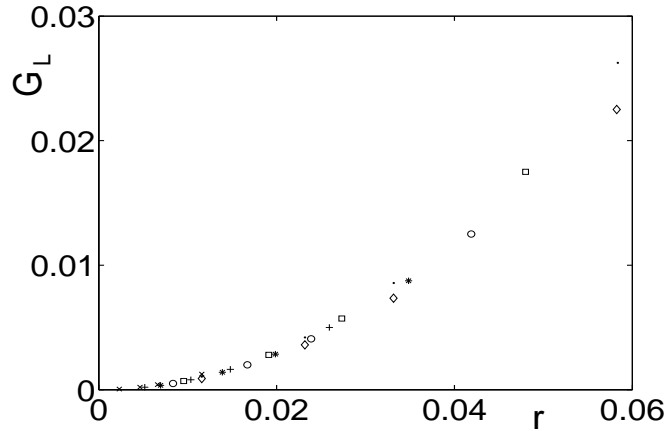}
\caption[a]{The distribution of zeroes for the bathroom-tile Ising
model with $L=M=N=40-200$ 
and  $j=1-4$ ($\times$), 
     $j=5-12$ ($+$),
     $j=13-16$ ($*$),
     $j=17-24$ ($\circ$),
     $j=25-32$ ($\put(1,0.5){\framebox(5,5)}~~$),
     $j=33-40$ ($\diamond$),
     $j=41-44$ ({\tiny{$\bullet$}}).
}
\label{Gbath}
\end{figure}

The physical ferromagnetic
critical point is given by 
$u =  (1/2) \left(\sqrt{4\sqrt{2} - 2} - \sqrt{2}\right) = 0.249\,038\,4\dots$,
corresponding to $\beta = 0.695\,074\,1\dots$ \cite{Moremore2D}.
In this region, the $j=1$ zeroes are four-fold degenerate, 
the $j=5$ are eight-fold degenerate, 
the $j=13$ zeroes are again four-fold,
the $j=17$, $j=25$ and $j=33$ zeroes are each eight-fold degenerate,
and the $j=41$ zeroes are four-fold degenerate.
The cumulative density of zeroes near 
this  ferromagnetic
critical point 
is depicted in Fig.~\ref{Gbath} for
$L=40,70,100$, and $200$ with  $j=1$--$44$ (seven data points for 
each $L$).
A three-parameter fit to the form (\ref{2nd}) clearly
shows that the curve goes through the origin. Indeed, such a fit 
to the above data  gives 
$a_3 = 0.000\,000\,7(830)$. Now, setting $a_3 =0$,
a two-parameter fit to the data yields
$a_2=1.998(18)$, corresponding to $\alpha = 0.002(18)$,
fully consistent with zero, as expected.

The physical antiferromagnetic
critical point is given by 
$u = 4.015\,445\,4\dots$, near which the zeroes again
have a one-dimensional
locus (as evident in Fig.~\ref{zeroes_bath40}).
The degeneracy pattern for the first 44 zeroes is the same as in the above
ferromagnetic critical case. A three-parameter fit yields $a_3 = 0.000\,01(3)$, and a 
two parameter fit to this data gives $a_2=2.03(2)$. Restricting
the fit closer to the origin by using the $j=1-16$ (3 data points for each $L$)
yields $a_2=1.9994(163)$, compatible with $\alpha = 0$.

The accumulation point between the ferromagnetic and unphysical
regions occurs at $u = - 0.601\,231\,8\dots$ (for which there is no
real $\beta$). Here the degeneracy pattern is  different to 
those above, with
the $j=1$ zeroes being four-fold degenerate, 
the $j=5$ zeroes eight-fold, 
the $j=13$ zeroes again four-fold,
the $j=17$ and  $j=25$ zeroes  each eight-fold degenerate
while the $j=33$ zeroes are four-fold and the
$j=37$ zeroes are eight-fold degenerate.
The density analysis again reveals a transition
($a_3=0$), with $a_2=2$ (e.g., the first 24 zeroes for $L=40-200$
give $a_2=1.993(12)$, corresponding to $\alpha = 0.007(12)$).

A similar accumulation pattern occurs at the boundary between the 
antiferromagnetic and unphysical O$_1$ phases
at $u=-1.663\,251\,9$, with the corresponding
density analysis yielding $a_2=2.0095(123)$.

At each of the above four
accumulation points, traditional FSS yields $\nu=1$ and $\lambda=2$.

\subsection{Wilson Fermions}
The partition function, $Z_L(\kappa)$ for a system of free Wilson fermions involves
an integral over Grassmann variables, which, on completion, leads to 
the determinant of the Wilson matrix, $M^{\rm{(0)}}$. Here $\kappa= 1/(2m_0+d)$ 
is the hopping parameter, $m_0$ is the dimensionless bare fermion mass and
$d$ is the lattice dimensionality (which is $2$ in our case).
It is well known that this system exhibits a phase transition
at $1/2\kappa = d = 2$, where massless fermions appear in the continuum 
limit \cite{MM}.
This determinant may be expressed as 
a product of eigenvalues, and, for even lattice extent, $L$,
\begin{equation}
Z_L(\kappa)  = 
 {\rm{det}} M^{(0)}= 
\prod_{\alpha=1}^2\prod_{p} \lambda_\alpha^{\rm{(0)}}(p)
\quad,
\end{equation}
where
\begin{equation}
\lambda_\alpha^{\rm{(0)}}(p)
=
\frac{1}{2\kappa}
- \sum_{\mu =1}^2\cos{p_\mu}
 + i (-1)^\alpha
 \sqrt{\sum_{\mu =1}^2\sin^2{p_\mu}
      }
\quad ,
\end{equation}
with $p_\mu = 2 \pi \hat{p}_\mu/L$ and 
where $\hat{p}_1=-(L-1)/2, -(L-3)/2,\dots,(L-1)/2$, 
while $\hat{p}_2=-L/2, -L/2+1,\dots, L/2$. These values
comply with standard boundary requirements for Grassmann
variables, namely that they are periodic in the spatial ($1$-) 
direction and antiperiodic in the temporal ($2$-) one  \cite{MM}.

\begin{figure}[t]
\vspace{6cm}
\includegraphics{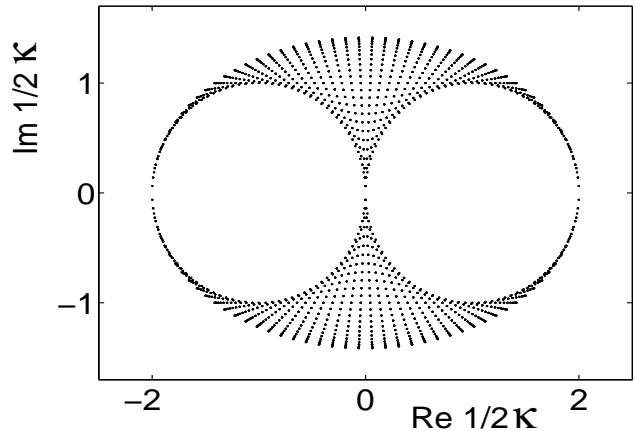}
\caption[a]{The partition function zeroes for the 
$L=50$ free Wilson fermions in the complex $1/2\kappa$ plane.}
\label{Wilson50}
\end{figure}

The complex hopping-parameter zeroes 
are easily and exactly extracted from the
multiplicative expression for the 
partition function (see \cite{KeSe02}) and
the zeroes for a system of size $L=50$ are depicted in Fig.~\ref{Wilson50}
 in the
complex $1/2\kappa$ plane.

A special feature of Wilson fermions is the occurence of so-called
doubler fermions. This means that apart from the physical 
critical point, which occurs where the zeroes accumulate at 
$1/2\kappa=2$ in the figure, there are lattice artefacts at $1/2\kappa=0$ and at 
$1/2\kappa=-2$ where further accumulations of zeroes, 
leading to critical behaviour, occur.

These Wilson-fermion zeroes clearly form a two-dimensional distribution.
They also come in degenerate sets, with the first and seventh 
zeroes being 2-fold degenerate, while the third and nineth are 
4-fold degenerate. So this system encapsulates both new features we seek to address. 

\begin{figure}[hbt]
\vspace{6cm}
\includegraphics{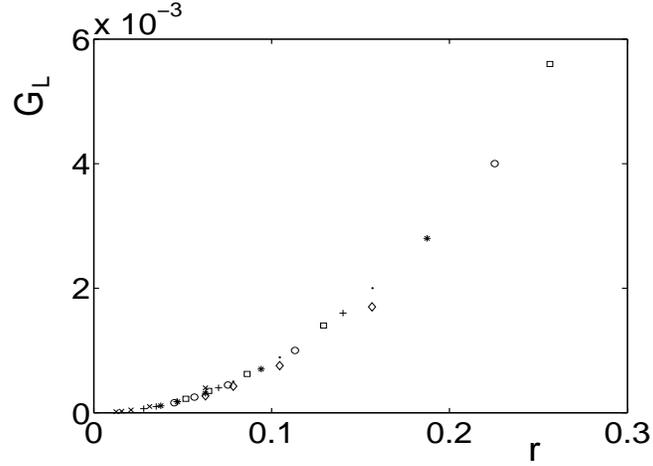}
\caption[a]{The distribution of zeroes near the physical critical point
$1/2\kappa = 2$
for free Wilson fermions
with $L=50-250$
and  $j=1-2$ ($\times$), 
     $j=3-6$ ($+$),
     $j=7-8$ ($*$),
     $j=9-12$ ($\circ$),
     $j=13-16$ ($\put(1,0.5){\framebox(5,5)}~~$),
     $j=17-18$ ($\diamond$),
     $j=19-23$ ({\tiny{$\bullet$}}).
}
\label{GWilson}
\end{figure}
The density plot for the zeroes near the physical transition is given in
Fig.~\ref{GWilson}.
Using the first twelve zeroes for lattices of size $L=50,100,150,200$, 
and $250$ (four data points for each lattice size), a three-parameter fit yields
$a_3 = 0.000\,005(29)$, convincing evidence that the density plot
indeed goes through the origin. The subsequent two-parameter fit yields
$a_2={1.996(11)}$, giving  $\alpha = 0$, as expected.

\begin{figure}[hbt]
\vspace{6cm}
\includegraphics{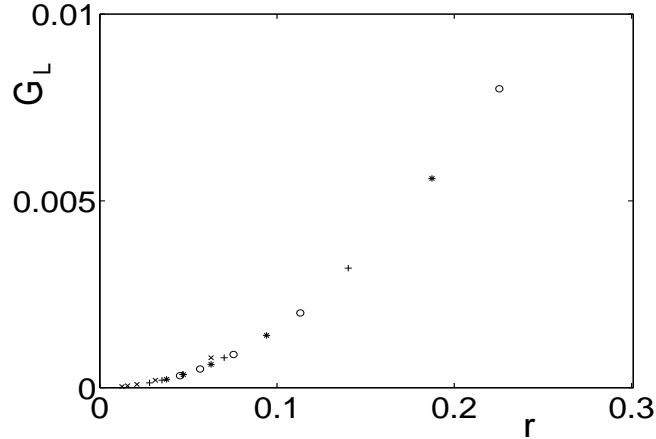}
\caption[a]{The distribution of zeroes near the artefactual critical point
$1/2\kappa = 0$
for free Wilson fermions
with $L=50-250$
and  $j=1-2$ ($\times$), 
     $j=3-6$ ($+$),
     $j=7-8$ ($*$),
     $j=9-12$ ($\circ$).
}
\label{Wilsonz0}
\end{figure}

It is worthwhile also applying  the method to the artifactual doubler transition
at $1/2\kappa=0$, 
where the two-dimensional nature of the distribution is more pronounced.
There, the density data again fall on a universal curve 
(see Fig.~\ref{Wilsonz0}) and $a_3$ is determined to be 
$0.000\,01(6)$. A two-parameter fit now yields 
$a_2={1.996(11)}$, again demonstrating that $\alpha$ is
zero and  the success of the method.
Finally, as in the other systems studied here,
traditional FSS yields $\nu=1$ and $\lambda=2$, so in each case
the shift exponent does not match the inverse of the
correlation-length exponent.
%

\section{Conclusions}
\setcounter{equation}{0}

A recently introduced technique to extract a continuous function,
in the form of the density of partition function zeroes, from sets 
of discrete data has been extended to deal with the general case
where ({\em i\/}) zeroes do not fall on a one-dimensional curve 
and/or where 
({\em ii\/})
multiple zeroes may occur. The technique is tested in a variety
of models which lie in the same universality class as the two-dimensional
Ising model and which exhibit various combinations of these general features. 
It is seen to be capable of direct determination of the strength
of the phase transition, as measured by the critical exponent 
$\alpha$. We have compared the results obtained from more standard 
finite-size scaling of the individual zeroes and also found good agreement.

It also perhaps worth highlighting that in this exercise we have found that formulating 
an Ising model with anisotropic couplings and Brascamp-Kunz boundary conditions
is straightforward and still leads to a simple product form for the finite lattice partition function,
a very useful property for investigating scaling.  
Though we have only touched on the topic briefly in this paper, the exotic critical points 
which appear at complex couplings in many models are
also amenable to our analysis, and we discuss this elsewhere.

\section{Acknowledgements}

W.J. and D.J. were partially supported by
EC IHP network
``Discrete Random Geometries: From Solid State Physics to Quantum Gravity''
{\it HPRN-CT-1999-000161}. 
RK would like to thank the Trinlat group at Trinity College
Dublin for hospitality during an extended visit.

\bigskip
%

\end{document}